\documentclass[]{aastex631}

\usepackage{amssymb}
\usepackage{amsmath}
\usepackage{graphicx}  
\usepackage{hyperref}

\usepackage{color}

\usepackage{soul}

\usepackage{url}
\usepackage{breakurl}
\usepackage{braket}

\begin{document}

\title{Signatures of an entangled graviton duet}

\correspondingauthor{Preston Jones}
\email{Preston.Jones1@erau.edu}

\author[0000-0001-9695-4367]{Preston Jones}

\author{Logan Finke}

\author{Joseph Ribaudo}

\affiliation{Embry-Riddle Aeronautical University \\
3700 Willow Creek Road \\
Prescott, AZ 86301, USA}

\date{\today}

\begin{abstract}
With the detection of the Higgs boson the Standard Model is an almost complete theory. Missing from the Model is a massless spin two boson, the graviton. We might reasonably expect that detection of gravitational radiation would include the observation of the graviton as the quanta of the radiation. However, the connection between gravitational radiation observations and the theoretical graviton has yet to be realized due to the physical limitations on single graviton detections. A more promising approach to demonstration of the non-classical nature of gravitational radiation, is to look at quantum entanglement in bipartite detections.
\end{abstract}






%

\section{Introduction}

Conspicuously absent from the standard model of particle physics is the spin-2 massless boson acting as the mediator particle for the gravitational force. The graviton has yet to be observed, either directly or indirectly, in any physical process that is in any way distinguishable from classical general relativity. The overwhelming majority of the effort, both theoretical and experimental, to demonstrate the existence of this hypothetical graviton is in finding an elusive projective measurement for a single graviton event. Any experimental validation of the existence of the graviton will require a meaningful theoretical framework for interpretation of the observations. Here the theoretical work has been consistent \citep{Dyson:2013hbl,2006FoPh...36.1801R,PhysRevD.109.044009}, repeatedly reaching the conclusion that no physically realizable experiment could ever identify the collapse of the graviton state-function in a projective measurement. The physical limitations on detections of the graviton can be understood by considering the exceedingly weak interaction of gravitons with the standard model fields. Indeed, as far back as 1975, Skoblev calculated the tree level diagrams for graviton-photon interactions  \citep{1975SvPhJ..18...62S}, with an astonishing cross-section of $\sigma \sim 10^{-110} ~ \rm{cm^2}$. So extraordinary as to put this among the smallest physically meaningful quantities, much smaller than even the square Planck length, $L^2_p \sim 10^{-66} ~ \rm{cm^2}$.

More recent and more promising theoretical work has accepted the unlikely or even physically impossible projective measurement of single gravitons. Instead this work considers measures of non-classicality associated with quantum entanglement \citep{2023IJMPA..3830005J,2025PhLB..86839628J,2025PRD.111d6004P,2025JHEP...08..104D}. The fortuitous separation distance of current gravitational wave detectors, e.g. $\sim 3,000 ~ \text{km}$ for Hanford-Livingston, is comparable to the wavelength of the hypothetical gravitons detected, e.g. $3,000 ~ \text{km}$ for graviton frequency of $100 \text{Hz}$, which makes experiments to identify entanglement measures in bipartite gravitational wave detections a reasonable possibility. This hope is bolstered by the low effective graviton number for current detector sensitivities and the prospect of single effective graviton detections per cycle for the next generation detectors.

\section{Bipartite quantum signatures}

\begin{figure}[htp!]
 \centering

    \includegraphics[width=0.75\linewidth]{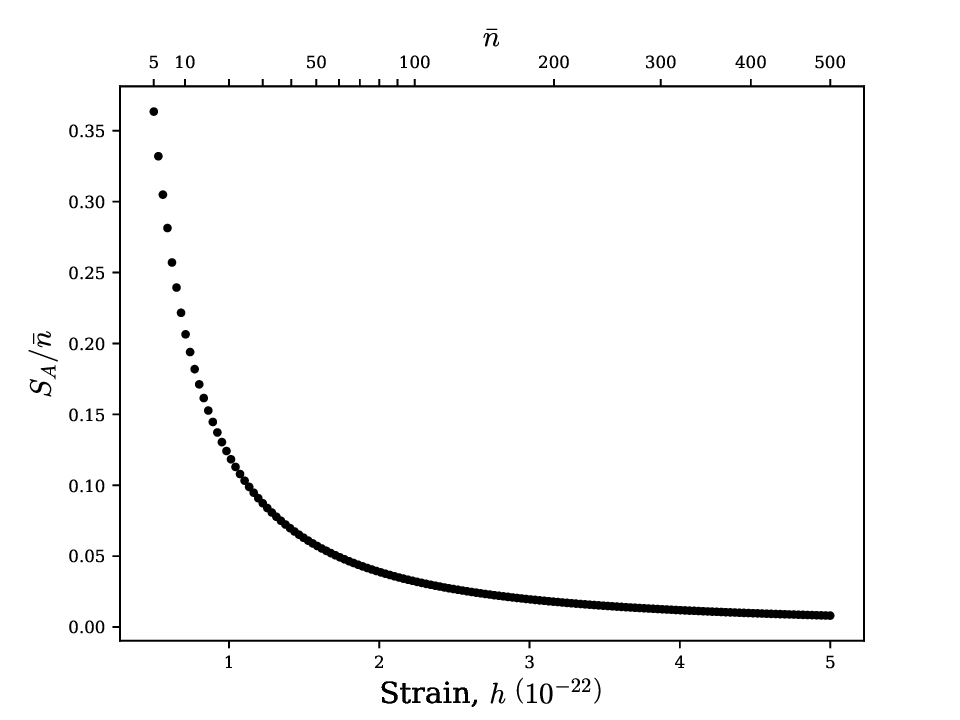}

 \caption{Entanglement entropy per mean number of gravitons $S_A/\bar{n}$ equation \eqref{EEntropySpectral} plotted against the detected gravitational wave strain amplitude $h$ and the mean graviton number per cycle $\bar{n}=\left(2 \times 10^{45}   \right) h^2$.}
 \label{MIentanglement}
\end{figure}

A fixed number representation of bipartite gravitational wave detections \citep{2025PhLB..86839628J} has already been shown to produce a meaningful measure of non-classicality through measurement-induced entanglement, Figure \ref{MIentanglement}. This is in stark contrast with efforts to develop measures of non-classicality based on coherent state representations \citep{PhysRevD.109.044009} and single-point detections. A coherent state is defined as an eigenstate of the annihilation operator, $\hat{a} \ket{\alpha}= \alpha  \ket{\alpha}$ as opposed to the ``lowering'' of the number state for a finite number state, $\hat{a} \ket{n}= \sqrt{n}  \ket{n-1}$. Unlike the fixed number state representation any quantum measure or signature based on coherent states would be ambiguous due to the optical equivalence theorem and the formal equivalence of classical and non-classical representations of the correlation functions \citep{Sudarshan:1963ts,1963PhRv..131.2766G}.


To develop a meaningful bipartite measure of non-classicality we focus our attention on fixed number states and measurement-induced entanglement of gravitational wave (GW) detections \citep{2025PhLB..86839628J}. Assuming equivalent detector subsystems $A$ and $B$ the density operator $\hat{\rho} = \ket{\varphi_{n}}  \bra{\varphi_{n}}$ for a bipartite pure state can be written as,

\begin{equation}
\hat{\rho} =   \sum_{j=0}^{n}  \sum_{k=0}^{n}   c \left(n, k \right) c^{*} \left(n, j \right) \ket{k}_{A}  \ket{n - k}_{B} \bra{n - j}_{B}  \bra{j}_{A} ~.
\label{bipartiteDMgen}
\end{equation}

\noindent The entanglement entropy for subsystem $A$ is then calculated taking the partial trace $\hat{\rho}_A = {\rm tr}_B \left( \hat{\rho} \right)$,

\begin{equation}
S_A  = - {\rm tr} \left( \hat{\rho}_A  {\rm ln} ~ \hat{\rho}_A \right) = -  \sum_{k=0}^{n}  \lambda \left( k \right) {\rm ln} \left(  \lambda \left( k \right) \right) ~,
\label{EEntropySpectral}
\end{equation}

\noindent with spectral decomposition, $\lambda \left( k \right) = c \left(n, k \right) c^{*} \left(n, k \right) $. A symmetric probability distribution can be constructed \citep{2025PhLB..86839628J} using Gaussian amplitudes,

\begin{equation}
\ket{\varphi_{n}} = \mathcal{N} \left( n , \bar{n} \right)  \sum_{k=0}^{n}  e^{ - \frac{ \left(k - \bar{n} \right)^2}{4 \sigma^2}} e^{i \phi_k}  \ket{k}_{A}  \ket{n - k}_{B} ~,
\label{Gbipartitefixed}
\end{equation}

\noindent following \citep{2025PhLB..86839628J} we take $\sigma^2=\bar{n}$. After normalization the elements of the spectral decomposition are found to be, 

\begin{equation}
\lambda \left( k \right) =  \left( \sum_{j=0}^{n}   e^{ - \frac{ \left(j - \bar{n} \right)^2}{2  \bar{n}}} \right)^{-1} e^{ - \frac{ \left(k - \bar{n} \right)^2}{2  \bar{n}}}  ~,
\label{SpecDecomG}
\end{equation}

\noindent substituting into equation \eqref{EEntropySpectral} the entanglement entropy is plotted in Figure \ref{MIentanglement} as a fraction of mean graviton number. Figure \ref{MIentanglement} is similar to \citep{2025PhLB..86839628J}, since both plots are based on \eqref{EEntropySpectral}, however the range for strain is lower and we include the effective graviton number $\bar{n}$ to better represent the entanglement for low $\bar{n}$. The effective mean graviton number is calculated by taking the average energy for an interferometer gravitational wave detector arm, characterized \citep{2018CQGra..35sLT02L,2018PRD..98l4006P,2023IJMPA..3830005J,2025PhLB..86839628J} by the mirror deviation $\xi$ from equilibrium  $ E =  \frac{1}{2} M \dot{\xi}^2 = \frac{1}{8} M L^2 \omega^2  h^2$ which we double for two detector arms. The mean graviton number per cycle is then,

 \begin{equation} 
 \bar{n} =\frac{  \frac{1}{4} M L^2 \omega^2 h^2 }{ \hbar \omega } =   \frac{1}{4 \hbar}   M L^2 \omega  h^2.
 \label{gravitonNumber}
\end{equation}

\noindent Taking the mirror masses $M = 50 ~ \rm{kg} $, arm lengths $L = 4,000 ~ \rm{m} $, and angular frequency $\omega = 10^3 ~  \rm{\frac{rad}{s}}$ the mean detector graviton number $ \bar{n}  = \left(2 \times 10^{45}   \right) h^2 $. Which can be inverted to calculate the strain amplitude for $\bar{n}  = 1$ associated with a single effective graviton detection per cycle, $h \sim 10^{-23}$. Our calculation of the effective graviton number follows the literature \citep{2018CQGra..35sLT02L,2023IJMPA..3830005J,2025PhLB..86839628J} in assuming one mirror per arm of the interferometers. Including the input mirrors, as well as the end mirrors, would double the graviton number calculations and should be considered in future work. With increasing detector sensitivity and the effective graviton number approaching one we would anticipate highly non-classical anti-correlations comparable to single photon interference \citep{Grangier_1986} in future detectors, e.g. the Einstein Telescope \citep{Kaiser21}. 


\section{Two are better than one}

The growth in the creative landscape of schema to identify signatures of non-classicality in single-point detections of gravitons shows little sign of abating. However imaginative these efforts might be, any scheme for detection of single gravitons must directly address the simple and compelling arguments by Dyson \citep{Dyson:2013hbl} challenging the feasibility of detecting single gravitons. Dyson examines three types of detectors and concludes that these detectors would fail to detect gravitons. Notably, LIGO type detectors are shown to fail based on the fundamental laws of physics. Gravito-optical or gravito-electric detectors would fail due to unavoidable background neutrino noise in the real Universe. Additionally, any method based on graviton-conversion would not be possible due to the size limitations for physically possible detectors. Dyson also considers a fourth type of detector proposed by Guth \citep{Guth:1997wk} to identify single graviton processes in the early Universe; however, the plausibility of this scheme over the other three has not yet been demonstrated. We should also bring attention to another detector scheme \citep{2014NJPh...16h5003S,2024NatCo..15.7229T} that proposes the use of resonance detectors for signal amplification. While these detector proposals were not directly addressed by Dyson, the new schemes have not fully addressed the restrictions of operating in the real Universe.

Recognizing the pessimistic results for the development of single-point graviton detections we turn our attention to bipartite detections.  One of the most promising bipartite methods for the demonstration of the existence of the graviton or non-classicality in gravitational waves are intensity correlations associated with Hanbury Brown and Twiss intensity (HBT) interferometry. While this approach to searching for the graviton has received some attention \citep{2019PRD..99h4010K}, HBT interferometry does not always make a clear distinction between classical and non-classical correlations.  The optical equivalence theorem \citep{Sudarshan:1963ts,1963PhRv..131.2766G} can be extended to a gravito-optical equivalence by assuming a single polarization for the gravitational radiation, making it challenging to distinguish between classical and non-classical correlations. The theorem introduces a formal equality between the classical and non-classical correlation functions for GW detections, which can be expressed succinctly \citep{2023IJMPA..3830005J} as ${G}^{(2)}_{c} \left(x_1,x_2 \right) = \mathcal{G}^{(2)} \left(x_1,x_2 \right)$ where the classical and non-classical unweighted second order correlation functions for GW detections are equivalent. This formal equality is based on the representation of the detected signal as coherent states, which are an ``overcomplete set of eigenstates of the destruction operator'' \citep{Sudarshan:1963ts}.

A meaningful measure of non-classicality for GW detections has already been realized through measurement-induced entanglement \citep{2025PhLB..86839628J}, as illustrated in Figure \ref{MIentanglement}, which is similar to measurement-induced entanglement realized in other systems \citep{PhysRevX.15.021059,2023Natur.622..481G}. The measurement-induced entanglement is calculated from fixed number states and avoids the ambiguity associated with coherent states and the optical equivalence theorem. Figure \ref{MIentanglement} evinces the potential of bipartite detections in demonstrating non-classicality in GW detections. A clear distinction needs to be made here between a measure as a theoretical calculation and a measurement as a resolved detector response. While several theoretical signatures of non-classicality have been developed for bipartite detections, e.g. entanglement entropy \citep{2025PhLB..86839628J} and sub-Poissonian correlations from squeezed states \citep{2019PRD..99h4010K}, no scheme has been developed to connect the entanglement entropy as a measure to a concrete measurement of the detector response. It does illustrate the value of exploring measurement schemes for the GW detectors operating as HBT interferometers. Progress \citep{2023IJMPA..3830005J} has been made in developing a meaningful measure through the calculation of intensity correlations for existing GW detections. Bipartite detections are an encouraging approach to the demonstration of non-classicality in GW detections, and deserve much greater attention from the theoretical physics community.

There are a great many possibilities for GW detection measurements based on the characterization of the two point GW detections as HBT interferometers \citep{2023IJMPA..3830005J}. One such possibility is to develop measurement schemes which measure detector coincidence rates and identify increased detection coincidence due to bosonic bunching. The measurement of coincidence rates has been shown to be a useful measure of non-classicality \citep{Jeltes07} by comparing bunching and anti-bunching for bosonic and fermionic helium falling onto a detector plate. The freedom in this measurement scheme to compare bosonic and fermionic detector responses illuminates perhaps one of the most interesting challenges in the hunt for the graviton. Detectable gravitational radiation is not created in the laboratory and we must be very clever in both the construction of meaningful measurements of non-classicality and in the interpretation of those measurements.

In quantum optics, sub-Poissonian correlations have been established as an unambiguous measure of non-classicality and sub-Poissonian correlations are produced by measurement of squeezed states. Squeezed states for gravitational radiation \citep{2019PRD..99h4010K} should produce the same unambiguous measure of non-classicality in bipartite GW detections. This does establish a potential bipartite measure of non-classicality that would also be measurable. Demonstrating, at least in principle, that resolvable bipartite measures of non-classicality are possible. However, there is no established source for production of squeezed states in GWs. The best path to the identification of non-classicality in GW detections is to focus on GW signals that current and near future detectors can observe. The effective graviton number $\bar{n}$ is proportional to the detector response and for resolvable gravitational wave detections, we can see in Figure \ref{MIentanglement} that non-classical contributions to bipartite correlations, e.g. from HBT interferometry, for gravitational wave detections should be appreciable when approaching $\bar{n} \sim 10$. While bipartite detection schemes do not yet offer both practical measures and measurements of non-classicality, they are arguably the most promising approach to finding the graviton. 




\section{A few final remarks}

Considering the pessimistic prospects for measures of non-classicality involving single-point detections, it is prudent to redirect our efforts to bipartite detections schemes. Here the projective measurement of single gravitons and the associated physical restrictions on the detectors is replaced by non-classical detection interactions. Several promising measures of non-classicality for GW detections have already been proposed, and future prospects appear limited only by the imagination of the theoretical physics community.

We should also point out that the development of bipartite measures of non-classicality for GW detections is an interesting problem in the foundations of quantum physics. The non-classicality of the standard model fields can generally be demonstrated by measurement of single quanta events. This measure of non-classicality is conceptually simple and the detector response is unambiguous. The gravitational field on the other hand is not known a priori to be quantized and the existence of the graviton is not certain. Other measures of non-classicality are needed in the search for the graviton which must be understood at the most fundamental level of quantum physics. There has been only a handful of papers on bipartite measures of non-classicality in gravitational wave detections and a great many papers on single point detections. One of our goals in this paper is to encourage greater interest in the promise of searching for the graviton through the development of bipartite measures of non-classicality. 

In closing we note that the problem of finding the graviton through bipartite detections is very much a problem in the increasingly important field of quantum information theory.


\bibliography{References.bib}{}

@ARTICLE{2025PhLB..86839628J,
       author = {{Jones}, Preston and {Bailey}, Quentin G. and {Gretarsson}, Andri and {Poon}, Edward},
        title = "{Measurement-induced entanglement entropy of gravitational wave detections}",
      journal = {Physics Letters B},
         year = 2025,
        month = sep,
       volume = {868},
          eid = {139628},
        pages = {139628},
          doi = {10.1016/j.physletb.2025.139628},
archivePrefix = {arXiv},
       eprint = {2411.15632},
 primaryClass = {gr-qc},
       adsurl = {https://ui.adsabs.harvard.edu/abs/2025PhLB..86839628J}
}

@article{Dyson:2013hbl,
    author = {Dyson, Freeman},
    title = {Is a graviton detectable?},
    doi = {10.1142/S0217751X1330041X},
    journal = {International Journal of Modern
Physics A},
    volume = {28},
    pages = {1330041},
    year = {2013}
}

@ARTICLE{Jeltes07,
       author = {{Jeltes}, Tom and {Mcnamara}, John M. and {Hogervorst}, Wim and {Vassen}, Wim and {Krachmalnicoff}, Valentina and {Schellekens}, Martijn and {Perrin}, Aur{\'e}lien and {Chang}, Hong and {Boiron}, Denis and {Aspect}, Alain and {Westbrook}, Christoph I.},
        title = "{Hanbury Brown Twiss effect for bosons versus fermions}",
      journal = {Nature},
    volume = {445},
         year = 2007,
        month = {January},
          eid = {cond-mat/0612278},
        pages = {402-405},
          doi = {https://doi.org/10.1038/nature05513},
archivePrefix = {arXiv},
       eprint = {cond-mat/0612278},
 primaryClass = {cond-mat.other},
       adsurl = {https://ui.adsabs.harvard.edu/abs/2006cond.mat.12278J}
}

@ARTICLE{1975SvPhJ..18...62S,
       author = {{Skobelev}, V.~V.},
        title = "{Graviton-photon interaction}",
      journal = {Soviet Physics Journal},
         year = 1975,
        month = jan,
       volume = {18},
       number = {1},
        pages = {62-65},
          doi = {10.1007/BF00889810},
       adsurl = {https://ui.adsabs.harvard.edu/abs/1975SvPhJ..18...62S}
}

@article{PhysRevD.109.044009,
  title = {Graviton detection and the quantization of gravity},
  author = {Carney, Daniel and Domcke, Valerie and Rodd, Nicholas L.},
  journal = {Phys. Rev. D},
  volume = {109},
  issue = {4},
  pages = {044009},
  numpages = {19},
  year = {2024},
  month = {Feb},
  publisher = {American Physical Society},
  doi = {10.1103/PhysRevD.109.044009},
  url = {https://link.aps.org/doi/10.1103/PhysRevD.109.044009}
}

@ARTICLE{2024NatCo..15.7229T,
       author = {{Tobar}, Germain and {Manikandan}, Sreenath K. and {Beitel}, Thomas and {Pikovski}, Igor},
        title = "{Detecting single gravitons with quantum sensing}",
      journal = {Nature Communications},
         year = 2024,
        month = dec,
       volume = {15},
       number = {1},
          eid = {7229},
        pages = {7229},
          doi = {10.1038/s41467-024-51420-8},
archivePrefix = {arXiv},
       eprint = {2308.15440},
 primaryClass = {quant-ph},
       adsurl = {https://ui.adsabs.harvard.edu/abs/2024NatCo..15.7229T}
}

@ARTICLE{2025JHEP...08..104D,
       author = {{Dutta}, Mainak and {Nandi}, Partha and {Majhi}, Bibhas Ranjan},
        title = "{Gravitationally induced entanglement at finite temperature: A memory-driven time-crystalline phase?}",
      journal = {JHEP},
         year = 2025,
        month = aug,
       volume = {2025},
       number = {8},
          eid = {104},
        pages = {104},
          doi = {10.1007/JHEP08(2025)104},
archivePrefix = {arXiv},
       eprint = {2503.19688},
 primaryClass = {gr-qc},
       adsurl = {https://ui.adsabs.harvard.edu/abs/2025JHEP...08..104D}
}

@ARTICLE{2025PRD.111d6004P,
       author = {{Parikh}, Maulik and {Setti}, Francesco},
        title = "{Quantum-gravitational noise correlation in nearby detectors}",
      journal = {Phys. Rev. D},
         year = 2025,
        month = feb,
       volume = {111},
       number = {4},
          eid = {046004},
        pages = {046004},
          doi = {10.1103/PhysRevD.111.046004},
archivePrefix = {arXiv},
       eprint = {2312.17335},
 primaryClass = {gr-qc},
       adsurl = {https://ui.adsabs.harvard.edu/abs/2025PRD.111d6004P}
}

@ARTICLE{2023IJMPA..3830005J,
       author = {{Jones}, Preston and {Barrett}, Alexander and {Carpenter}, Justin and {Gretarsson}, Andri and {Gretarsson}, Ellie and {Hughey}, Brennan and {Smith}, Darrel and {Zanolin}, Michele and {Singleton}, Douglas},
        title = "{Gravito-optics and intensity correlations for binary inspiral signal detections}",
      journal = {International Journal of Modern Physics A},
         year = 2023,
        month = mar,
       volume = {38},
          eid = {2330005-110},
        pages = {2330005-110},
          doi = {10.1142/S0217751X23300053},
archivePrefix = {arXiv},
       eprint = {1907.00100},
 primaryClass = {gr-qc},
       adsurl = {https://ui.adsabs.harvard.edu/abs/2023IJMPA..3830005J}
}

@ARTICLE{2019PRD..99h4010K,
       author = {{Kanno}, Sugumi and {Soda}, Jiro},
        title = "{Detecting nonclassical primordial gravitational waves with Hanbury-Brown-Twiss interferometry}",
      journal = {Phys. Rev. D},
         year = 2019,
        month = apr,
       volume = {99},
       number = {8},
          eid = {084010},
        pages = {084010},
          doi = {10.1103/PhysRevD.99.084010},
archivePrefix = {arXiv},
       eprint = {1810.07604},
 primaryClass = {hep-th},
       adsurl = {https://ui.adsabs.harvard.edu/abs/2019PRD..99h4010K}
}

@article{Sudarshan:1963ts,
    author = "Sudarshan, E. C. G.",
    title = "{Equivalence of semiclassical and quantum mechanical descriptions of statistical light beams}",
    doi = "10.1103/PhysRevLett.10.277",
    journal = "Phys. Rev. Lett.",
    volume = "10",
    pages = "277--279",
    year = "1963"
}

@ARTICLE{2006FoPh...36.1801R,
       author = {{Rothman}, Tony and {Boughn}, Stephen},
        title = "{Can Gravitons be Detected?}",
      journal = {Foundations of Physics},
         year = 2006,
        month = dec,
       volume = {36},
       number = {12},
        pages = {1801-1825},
          doi = {10.1007/s10701-006-9081-9},
archivePrefix = {arXiv},
       eprint = {gr-qc/0601043},
 primaryClass = {gr-qc},
       adsurl = {https://ui.adsabs.harvard.edu/abs/2006FoPh...36.1801R}
}

@ARTICLE{1963PhRv..131.2766G,
       author = {{Glauber}, Roy J.},
        title = "{Coherent and Incoherent States of the Radiation Field}",
      journal = {Physical Review},
         year = 1963,
        month = sep,
       volume = {131},
       number = {6},
        pages = {2766-2788},
          doi = {10.1103/PhysRev.131.2766},
       adsurl = {https://ui.adsabs.harvard.edu/abs/1963PhRv..131.2766G}
}

@book{Guth:1997wk,
    author = "Guth, Alan H.",
    title = "{The inflationary universe: The quest for a new theory of cosmic origins}",
  publisher = {Addison-Wesley},
  year = "1997"
}

@ARTICLE{2014NJPh...16h5003S,
       author = {{Sab{\'\i}n}, Carlos and {Bruschi}, David Edward and {Ahmadi}, Mehdi and {Fuentes}, Ivette},
        title = "{Phonon creation by gravitational waves}",
      journal = {New Journal of Physics},
         year = 2014,
        month = aug,
       volume = {16},
       number = {8},
          eid = {085003},
        pages = {085003},
          doi = {10.1088/1367-2630/16/8/085003},
archivePrefix = {arXiv},
       eprint = {1402.7009},
 primaryClass = {quant-ph},
       adsurl = {https://ui.adsabs.harvard.edu/abs/2014NJPh...16h5003S}
}

@ARTICLE{2018CQGra..35sLT02L,
       author = {{Lieu}, Richard},
        title = "{Exclusion of standard {\ensuremath{\hbar}}{\ensuremath{\omega}} gravitons by LIGO observation}",
      journal = {Classical and Quantum Gravity},
         year = 2018,
        month = oct,
       volume = {35},
       number = {19},
          eid = {19LT02},
        pages = {19LT02},
          doi = {10.1088/1361-6382/aadb30},
archivePrefix = {arXiv},
       eprint = {1712.04437},
 primaryClass = {astro-ph.HE},
       adsurl = {https://ui.adsabs.harvard.edu/abs/2018CQGra..35sLT02L}
}

@ARTICLE{2018PRD..98l4006P,
       author = {{Pang}, Belinda and {Chen}, Yanbei},
        title = "{Quantum interactions between a laser interferometer and gravitational waves}",
      journal = {Phys. Rev. D},
         year = 2018,
        month = dec,
       volume = {98},
       number = {12},
          eid = {124006},
        pages = {124006},
          doi = {10.1103/PhysRevD.98.124006},
archivePrefix = {arXiv},
       eprint = {1808.09122},
 primaryClass = {quant-ph},
       adsurl = {https://ui.adsabs.harvard.edu/abs/2018PRD..98l4006P}
}

@ARTICLE{2023Natur.622..481G,
       author = {{Google Quantum AI} and {Collaborators}, J.~C., Hoke and {Ippoliti}, M. and {Rosenberg}, E. and {Abanin}, D. and {Acharya}, R. and {Andersen}, T.~I. and {Ansmann}, M. and {Arute}, F. and {Arya}, K. and {Asfaw}, A. and {Atalaya}, J. and {Bardin}, J.~C. and {Bengtsson}, A. and {Bortoli}, G. and {Bourassa}, A. and {Bovaird}, J. and {Brill}, L. and {Broughton}, M. and {Buckley}, B.~B. and {Buell}, D.~A. and {Burger}, T. and {Burkett}, B. and {Bushnell}, N. and {Chen}, Z. and {Chiaro}, B. and {Chik}, D. and {Cogan}, J. and {Collins}, R. and {Conner}, P. and {Courtney}, W. and {Crook}, A.~L. and {Curtin}, B. and {Dau}, A.~G. and {Debroy}, D.~M. and {Del Toro Barba}, A. and {Demura}, S. and {Di Paolo}, A. and {Drozdov}, I.~K. and {Dunsworth}, A. and {Eppens}, D. and {Erickson}, C. and {Farhi}, E. and {Fatemi}, R. and {Ferreira}, V.~S. and {Burgos}, L.~F. and {Forati}, E. and {Fowler}, A.~G. and {Foxen}, B. and {Giang}, W. and {Gidney}, C. and {Gilboa}, D. and {Giustina}, M. and {Gosula}, R. and {Gross}, J.~A. and {Habegger}, S. and {Hamilton}, M.~C. and {Hansen}, M. and {Harrigan}, M.~P. and {Harrington}, S.~D. and {Heu}, P. and {Hoffmann}, M.~R. and {Hong}, S. and {Huang}, T. and {Huff}, A. and {Huggins}, W.~J. and {Isakov}, S.~V. and {Iveland}, J. and {Jeffrey}, E. and {Jiang}, Z. and {Jones}, C. and {Juhas}, P. and {Kafri}, D. and {Kechedzhi}, K. and {Khattar}, T. and {Khezri}, M. and {Kieferov{\'a}}, M. and {Kim}, S. and {Kitaev}, A. and {Klimov}, P.~V. and {Klots}, A.~R. and {Korotkov}, A.~N. and {Kostritsa}, F. and {Kreikebaum}, J.~M. and {Landhuis}, D. and {Laptev}, P. and {Lau}, K.-M. and {Laws}, L. and {Lee}, J. and {Lee}, K.~W. and {Lensky}, Y.~D. and {Lester}, B.~J. and {Lill}, A.~T. and {Liu}, W. and {Locharla}, A. and {Martin}, O. and {McClean}, J.~R. and {McEwen}, M. and {Miao}, K.~C. and {Mieszala}, A. and {Montazeri}, S. and {Morvan}, A. and {Movassagh}, R. and {Mruczkiewicz}, W. and {Neeley}, M. and {Neill}, C. and {Nersisyan}, A. and {Newman}, M. and {Ng}, J.~H. and {Nguyen}, A. and {Nguyen}, M. and {Niu}, M.~Y. and {O'Brien}, T.~E. and {Omonije}, S. and {Opremcak}, A. and {Petukhov}, A. and {Potter}, R. and {Pryadko}, L.~P. and {Quintana}, C. and {Rocque}, C. and {Rubin}, N.~C. and {Saei}, N. and {Sank}, D. and {Sankaragomathi}, K. and {Satzinger}, K.~J. and {Schurkus}, H.~F. and {Schuster}, C. and {Shearn}, M.~J. and {Shorter}, A. and {Shutty}, N. and {Shvarts}, V. and {Skruzny}, J. and {Smith}, W.~C. and {Somma}, R. and {Sterling}, G. and {Strain}, D. and {Szalay}, M. and {Torres}, A. and {Vidal}, G. and {Villalonga}, B. and {Heidweiller}, C.~V. and {White}, T. and {Woo}, B.~W.~K. and {Xing}, C. and {Yao}, Z.~J. and {Yeh}, P. and {Yoo}, J. and {Young}, G. and {Zalcman}, A. and {Zhang}, Y. and {Zhu}, N. and {Zobrist}, N. and {Neven}, H. and {Babbush}, R. and {Bacon}, D. and {Boixo}, S. and {Hilton}, J. and {Lucero}, E. and {Megrant}, A. and {Kelly}, J. and {Chen}, Y. and {Smelyanskiy}, V. and {Mi}, X. and {Khemani}, V. and {Roushan}, P.},
        title = "{Measurement-induced entanglement and teleportation on a noisy quantum processor}",
      journal = {Nature},
         year = 2023,
        month = oct,
       volume = {622},
       number = {7983},
        pages = {481-486},
          doi = {10.1038/s41586-023-06505-7},
archivePrefix = {arXiv},
       eprint = {2303.04792},
 primaryClass = {quant-ph},
       adsurl = {https://ui.adsabs.harvard.edu/abs/2023Natur.622..481G}
}

@article{PhysRevX.15.021059,
  title = {Measurement-Induced Entanglement and Complexity in Random Constant-Depth 2D Quantum Circuits},
  author = {McGinley, Max and Ho, Wen Wei and Malz, Daniel},
  journal = {Phys. Rev. X},
  volume = {15},
  issue = {2},
  pages = {021059},
  numpages = {35},
  year = {2025},
  month = {May},
  publisher = {American Physical Society},
  doi = {10.1103/PhysRevX.15.021059},
  url = {https://link.aps.org/doi/10.1103/PhysRevX.15.021059}
}

@article{Grangier_1986,
doi = {10.1209/0295-5075/1/4/004},
url = {https://doi.org/10.1209/0295-5075/1/4/004},
year = {1986},
month = {feb},
publisher = {},
volume = {1},
number = {4},
pages = {173},
author = {P. Grangier and G. Roger and A. Aspect},
title = {Experimental Evidence for a Photon Anticorrelation Effect on a Beam Splitter: A New Light on Single-Photon Interferences},
journal = {Europhysics Letters}
}

@article{Kaiser21,
author = {Kaiser, A and McWilliams, Sean},
year = {2021},
month = {03},
pages = {},
title = {Sensitivity of present and future detectors across the black-hole binary gravitational wave spectrum},
volume = {38},
journal = {Classical and Quantum Gravity},
doi = {10.1088/1361-6382/abd4f6}
}

\end{document}